\documentclass[fleqn,twoside]{article}

\usepackage{espcrc2}
\usepackage{graphicx}
\usepackage{bm}

\newcommand{\Vp}{{\bm p}}
\newcommand{\Vq}{{\bm q}}

\title{
Comparison of electron-nucleus quasi-elastic cross sections using
spectral functions with $(e,e')$ data from 0.5 GeV to 1.5 GeV and
effects on neutrino quasi-elastic cross sections }
\author{Hiroki Nakamura \address{Department of Physics,
Waseda University, Tokyo 169-8555, Japan}
\thanks{naka@hep.phys.waseda.ac.jp},
Ryoichi Seki
\address{Department of Physics, \\ California State
                 University, Northridge, California 91330, USA}${}^{,}$
\address{W.K. Kellogg Radiation Laboratory, 106-38,\\
                 California
                 Institute of Technology, Pasadena, California 91125, USA}
\thanks{rseki@krl.caltech.edu},
Makoto Sakuda \address{Institute for Particles and
Nuclear Studies, KEK, Tsukuba, 305-0801, Japan}
\thanks{sakuda@fphy.hep.okayama-u.ac.jp}
\thanks{Present address: Department of Physics, Okayama University, Okayama, 700
-8530 Japan}
}

\begin{document}

\begin{abstract}
We carry out a relativistic calculation of the cross sections of
electron-${}^{16}$O and -$^{12}$C quasi-elastic scattering and
quasi-free $\Delta$ production and compare with the $(e,e')$ data
systematically in the wide energy range of 0.5 -- 1.5 GeV.  Using
the same formalism, we examine the $\nu_\mu$ quasi-elastic
scattering from ${}^{16}$O.  The model incorporating the nuclear
correlation
 effects agrees better with the electron-nucleus scattering data
than a uniform Fermi-Gas model. In the neutrino quasi-elastic
scattering, the nuclear correlation has an appreciable effect on
the cross section of high-energy scattered leptons, and it may
have an important
consequence in the neutrino oscillation measurements aiming at a few \%
precision.
\end{abstract}

\maketitle

\section {Introduction}

The field of neutrino physics is developing rapidly after
atmospheric neutrino oscillations and solar neutrino oscillations
have been established \cite{Kajita,Solar,KamLAND}. Recently,  SK
collaboration has found an evidence for the oscillatory signature
in atmospheric neutrinos, improving the determination of $\Delta
m^2$ \cite{SKATM}, and K2K experiment has confirmed the neutrino
oscillations of atmospheric neutrinos at the 99.99\% CL
\cite{K2K,Nakaya}. These neutrino experiments measure the energy
and angle of muons produced in neutrino-nucleus interactions and
obtain the incident neutrino energy that determines the neutrino
oscillations. K2K takes data at $E_{\nu}$=0.5--3 GeV region, 
and the recent L/E analysis of the SK atmospheric neutrinos is
based on the dataset mainly from 0.5 to 25 GeV.
JPARC and NuMI neutrino
experiments \cite{JPARC,Nova} propose to measure
$\nu_{\mu}\rightarrow \nu_e $ oscillations and to determine
$\Delta m^2$ at $1\%$ level of the precision and $\sin^2 2\theta
_{13}$ above 0.006, using a narrow-band neutrino beam at
$E_{\nu}=0.8$ GeV (JPARC) and 2.0 GeV (NuMI, off-axis).

It is thus vital that theoretical calculations of the cross
sections and spectra could be carried out with a similar
reliability. For this, both of neutrino-nucleus reactions and
relevant nuclear structure have to be well under control. At
$E_{\nu}$=3 GeV or less, quasi-elastic scattering and quasi-free
$\Delta$ production dominates the neutrino-nucleus reactions. The
reactions at this energy region are associated with a wide range
of the momentum transfer and thus involve various aspects of
nuclear structure. Because the momentum transfer involved is large
enough that nuclear correlations are important as the relevant
nuclear structure beyond a mean-field description. Furthermore,
the neutrino-nucleus reactions do occur, in fact rather
appreciably, with a small momentum transfer, and requires a
careful treatment of nuclear properties and reactions, as observed
in the deficit of events in the low $Q^2 <0.2\ ({\rm GeV}/c)^2$
region \cite{Nakaya,Ishida}, compared to what is expected based on
a Fermi gas model.

In this paper, we focus on the aspects of effects of nuclear
correlations.  Using the same formalism, we study both of electron
and neutrino quasi-elastic scattering in the energy range of 0.5
and 1.5 GeV. For numerical calculations, we mostly take ${}^{16}$O
as it is the main target nucleus in SK, K2K, and other
experiments. We calculate the cross sections of
electron-${}^{16}$O quasi-elastic scattering and quasi-free
$\Delta$ production, using a relativistic plane-wave-impulse
approximation (PWIA) formalism \cite{Smith,Seki} with a uniform
Fermi-gas model and a spectral function \cite{Benhar}, and compare
with $(e,e')$ data, so as to examine the validity of the
calculation. The PWIA neglects final-state interactions, but we
will estimate their effects using simple formalisms.  Following
the electron scattering, we examine neutrino-${}^{16}$O scattering
to see how the effects observed in the electron scattering
manifest in the neutrino scattering.

\section{Formalism}

Electron-nucleus quasi-elastic cross section is expressed as
\begin{eqnarray}
\lefteqn{
\frac{d\sigma}{dE\,  ' d\Omega}=
} \nonumber \\ & &
\frac{k'}{8(2\pi)^4M_AE}
\int d^3{\bm p}F({\bm p},{\bm q},\omega) \sum_{\rm spin}|{\cal M}_{e N}|^2,
\end{eqnarray}
where $E$ is the incident electron energy, $M_A$ the mass of the
target nucleus, $E'$ and $k'$ the energy and momentum of the
scattered electron, respectively. ${\cal M}_{eN}$ is the invariant
amplitude of electron-nucleon elastic scattering.
$F(\Vp,\Vq,\omega)$ is the nuclear effect contribution expressed
in terms of $\Vp$, $\Vq$ and $\omega$; the initial nucleon
momentum, the momentum transfer, and the energy transfer,
respectively. $F(\Vp,\Vq,\omega)$ is proportional to the imaginary
part of the $1p1h$ Green's function. For high-energy scattering,
we apply a factorization approximation, a convolution of the $1p$
and $1h$ Green's functions:
\begin{eqnarray}
\lefteqn{F({\bm p},{\bm q},\omega) =} \nonumber \\ & &
\frac{1}{2M_A}
   \int d\omega'P_h(\bm p,\omega')P_p(\bm p +\bm q,\omega-\omega').
\end{eqnarray}
Here, $P_h(\Vp,\omega)$ is proportional to the $1h$ Green's
function, and is proportional to the nuclear spectral function.
The spectral function describes the probability of removing a
nucleon of the momentum $\Vp$ with the removal energy $\omega$
from the nucleus, which is usually referred to as the (nuclear)
spectral function. $P_p(\Vp,\omega)$ is proportional to the $1h$
Green's function. It describes the probability for adding a
nucleon, thus describing the final-state interactions of the
knocked-out nucleon as it goes out the residual target nucleus.

For the neutrino-nucleus quasi-elastic cross section, the
invariant amplitude of electron-nucleon scattering is replaced by
that of neutrino-nucleon scattering \cite{ls} with the same
$F(\Vp,\Vq,\omega)$.  For the quasi-free $\Delta$ production cross
section, different forms are used for both of the invariant
amplitude and $P_p(\Vp,\omega)$.

The electron-nucleon scattering amplitudes used in this
calculation are of the standard form, but we use the most
up-to-date vector nucleon form factors \cite{brash,bosted} and
$N-\Delta$ transition form factors \cite{PaschosYu}. As to the
invariant neutrino-nucleon amplitude, we use the standard dipole
form factor with $M_A=1.07$ GeV.

\begin{figure}[tb]
\begin{center}
\includegraphics[scale=0.25]{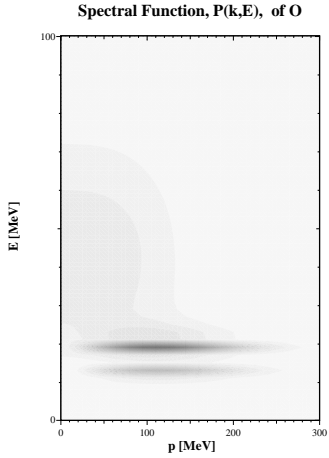}
\end{center}
\caption{Contour plot of spectral function $P(\bm p,E)$ of
${}^{16}$O \cite{benhar}.} \label{pke}
\end{figure}

The models used in this paper are the relativistic, simple
Fermi-gas model (FG) and a realistic spectral function model (SF).
FG was used by E. Moniz and his collaborators more than three
decades ago \cite{moniz,monizlett}.  It is known to yield a good
description of the total cross section, but to fail to provide a
reliable detailed description such the transverse and longitudinal
cross sections separately \cite{TL}. In FG, $P_h(\Vp,\omega)$ and
$P_p(\Vp,\omega)$ are given by
\begin{eqnarray}
P_h(\bm p,\omega) &=&\frac{1}{E_p}\theta(P_F-|\bm p|)\delta(E_p+\omega) \\
P_p(\bm p',\omega)
&=&\frac{1}{E_p'}\theta(|p'|-P_F)\delta(E_p'-\omega), \label{FGPp}
\end{eqnarray}
where $E_p$ is the effective Fermi-momentum parameter, $E_p =
\sqrt{p^2+M^2} -E_B$ with $E_B$ the effective nuclear
binding-energy parameter and $M$ the nucleon mass, and $E_p'=
\sqrt{{p'}^2+M^2}$. In our calculation, we take $P_F$ to be 225
MeV and $E_B$ 27 MeV \cite{monizlett}. Note that Eq. (\ref{FGPp})
describes the Pauli blocking condition in the final state.

SF uses a more realistic description of the initial, ground-state
nucleus. $P_h(\Vp,\omega)$ and $P_p(\Vp,\omega)$ in this model are
given by:
\begin{eqnarray}
P_h(\bm p,\omega) &=&\frac{1}{E_p}P(\bm p,\omega) \\
P_p(\bm p',\omega) &=&\frac{1}{E_p'}\delta(E_p'-\omega) \;,
\label{SFPp}
\end{eqnarray}
where $P(\bm p, \omega)$ is the spectral function.  In this work,
we use $P(\bm p, \omega)$ that was calculated by Benhar {\it et
al.} \cite{benhar}, which includes the nuclear shell structure
combined with the short-range correlations under a local density
approximation. Note that Eq. (\ref{SFPp}) for $P_p(\Vp,\omega)$
represents the PWIA we use.

\section{Numerical results}

\subsection{e-nucleus quasi-elastic scattering and quasi-free
$\Delta$ production}

\begin{figure*}[htb]
\begin{center}
\includegraphics[scale=0.5]{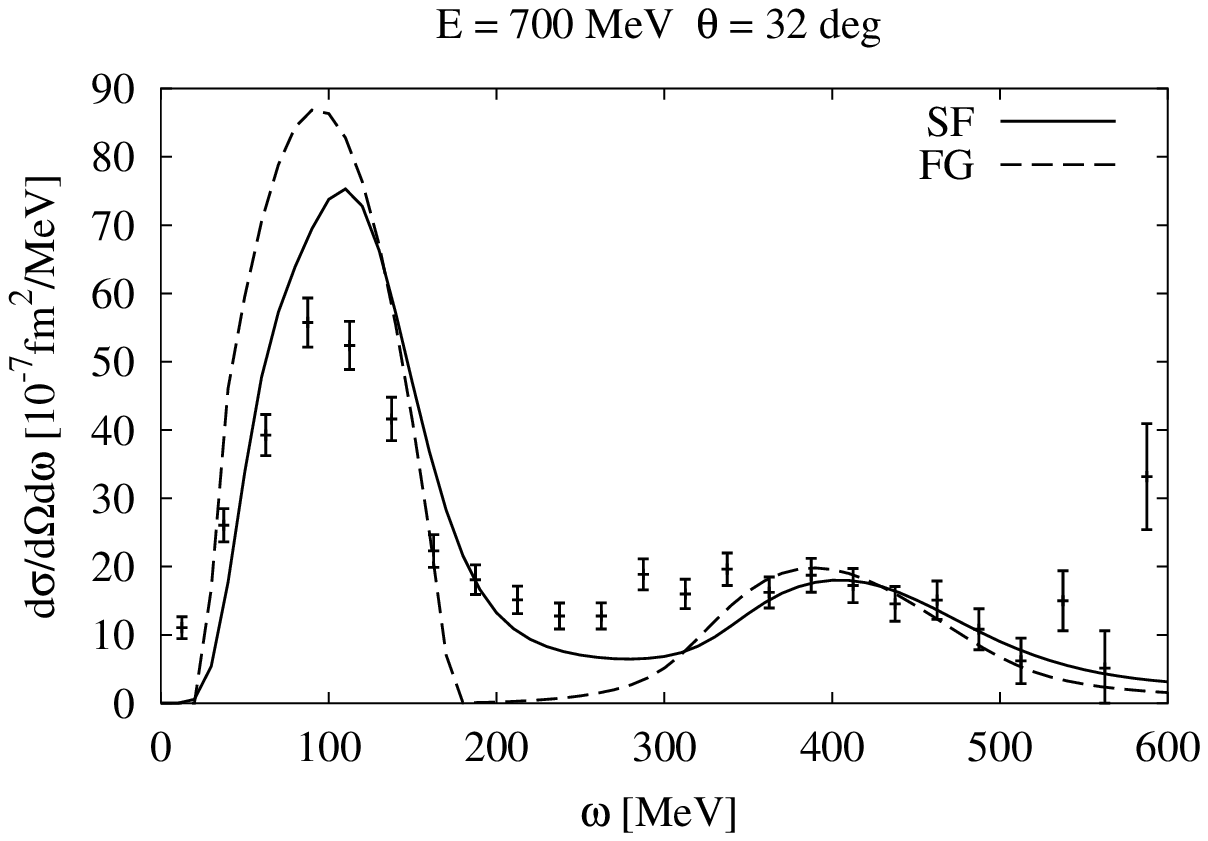}%
\includegraphics[scale=0.5]{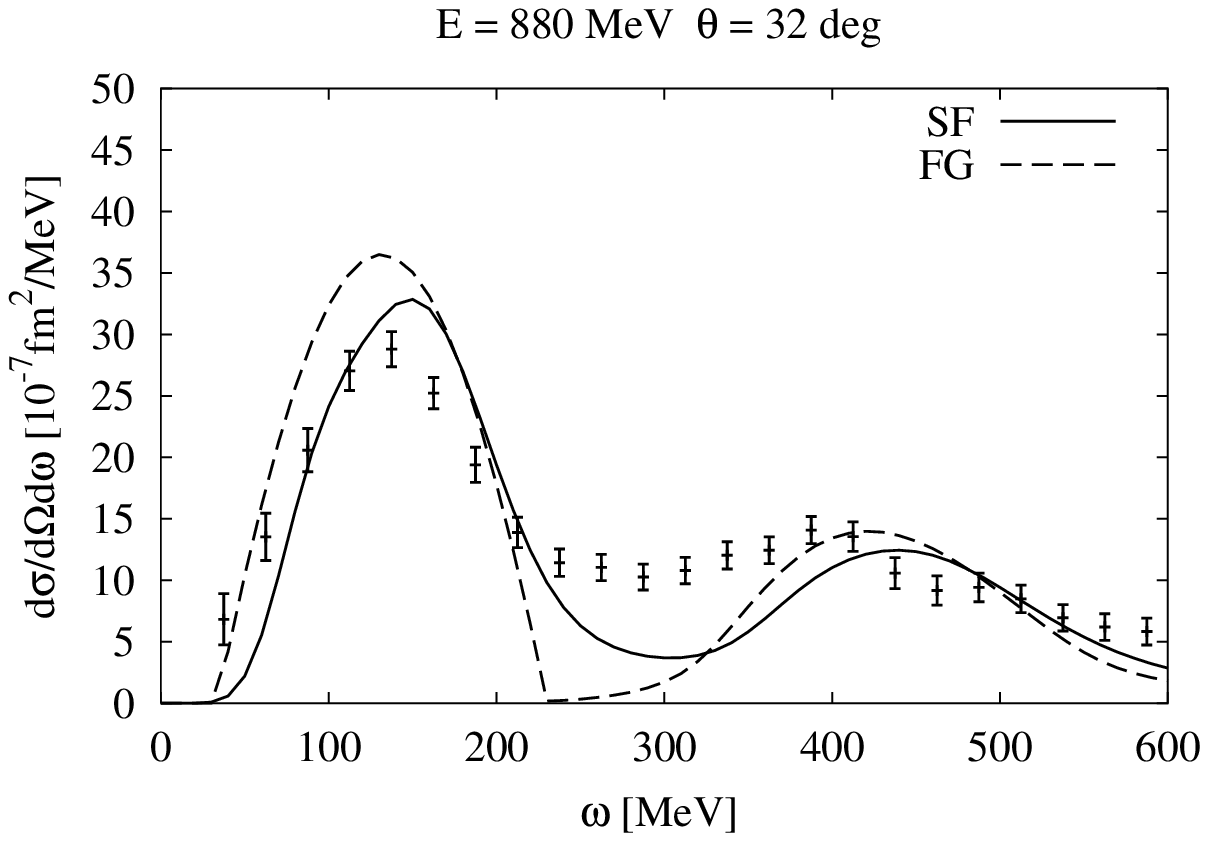}
\includegraphics[scale=0.5]{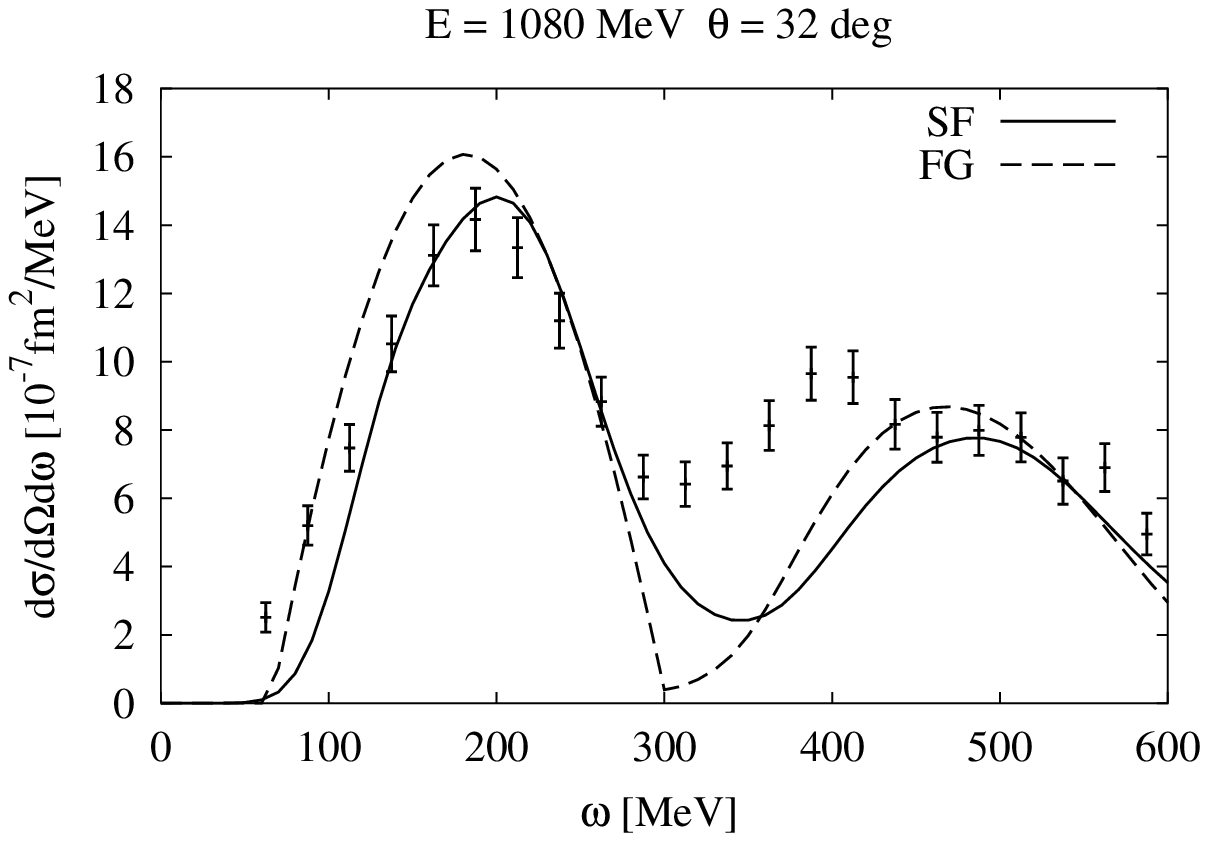}%
\includegraphics[scale=0.5]{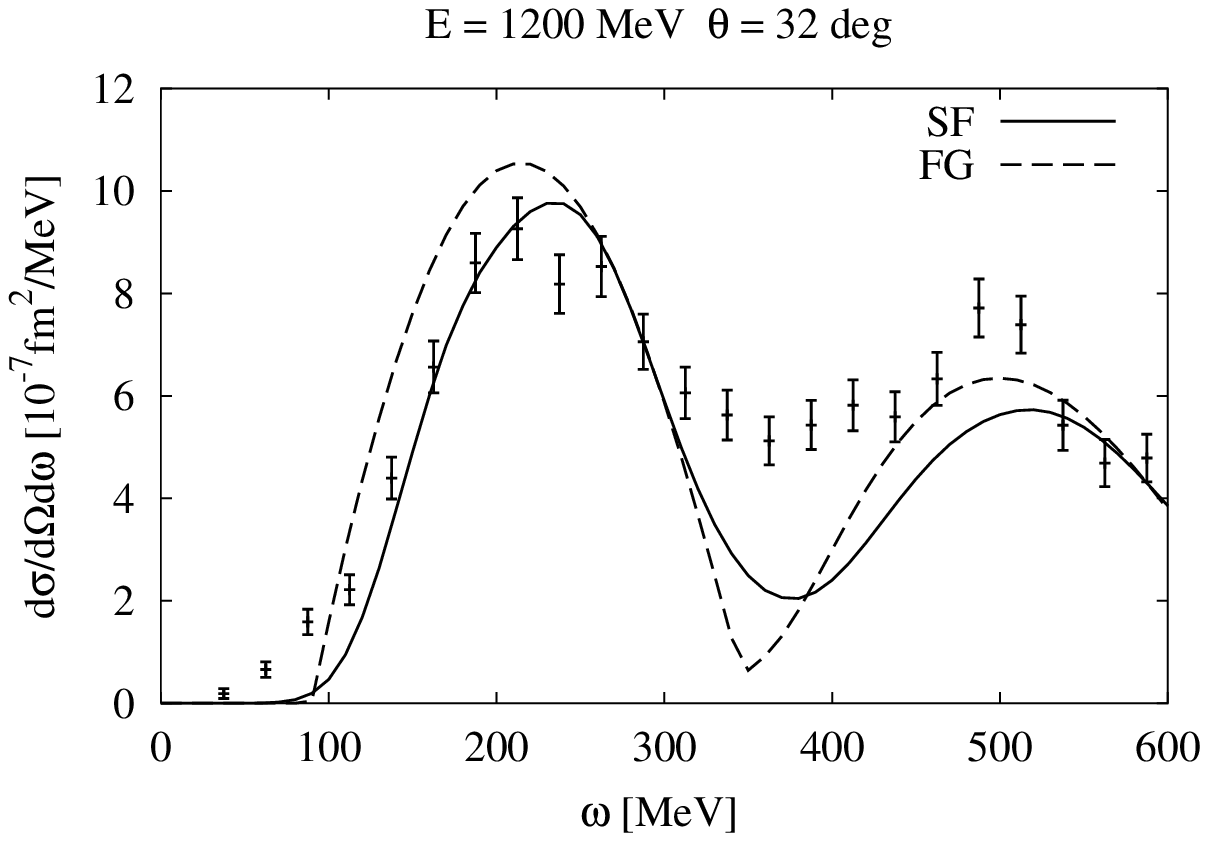}%
\end{center}
\caption{Combined cross sections of electron-${}^{16}$O
quasi-elastic scattering and quasi-free $\Delta$ resonance
production, using a spectral function (SF, solid curves) and a
simple Fermi gas model (FG, dash curves), compared with the
$(e,e')$ data \cite{anghinolfi}. } \label{oxygen}
\end{figure*}

We now compare our calculation with electron-${}^{16}$O scattering
data at various incident energies.  For a reference, we also
include the results for electron-${}^{12}$C quasi-elastic
scattering.  The experimental data we compare to are
${}^{16}$O$(e',e)$ at the scattering angle $\theta= 32^\circ$ for
the incident energy $E=$0.7, 0.88, 1.08, and 1.2 GeV
\cite{anghinolfi}, ${}^{12}$C$(e',e)$ at $\theta=60^\circ$ for
$E=0.5$ GeV \cite{whitney} and at $\theta=50.4^\circ$ for $E =
0.78$ GeV \cite{garino}.

Figure \ref{oxygen} includes four figures of the ${}^{16}$O cross
sections at the four incident energies. The quasi-elastic
scattering contribution appears as the most prominent peak on the
left-handed side of each figure.  We see that FG overestimates the
quasi-elastic contribution at all energies, but SF agrees with the
data better especially above 1 GeV. The FG quasi-elastic
contribution has a parabola shape, but the SF contribution has
tails at both sides of the energy, especially prominent on the
high energy side. In SF, the strength at the peak is reduced,
moving mostly to the high-energy tail region, as a consequence of
the inclusion of high-momentum components due to nuclear
correlations, in the spectral function in $F({\bm p},{\bm
q},\omega)$. FG and SF yield similar contributions to the
quasi-free $\Delta$ peak, in a good agreement with the data at all
energies. In the ``dip'' region between the two peaks, both of the
FG and SF contributions are smaller than the data.  SF gives,
however, larger contributions than FG and thus a better agreement
with the data, again as a consequence of the inclusion of nuclear
correlations in $F({\bm p},{\bm q},\omega)$.

\begin{figure}[htb]
\begin{center}
\includegraphics[scale=0.5]{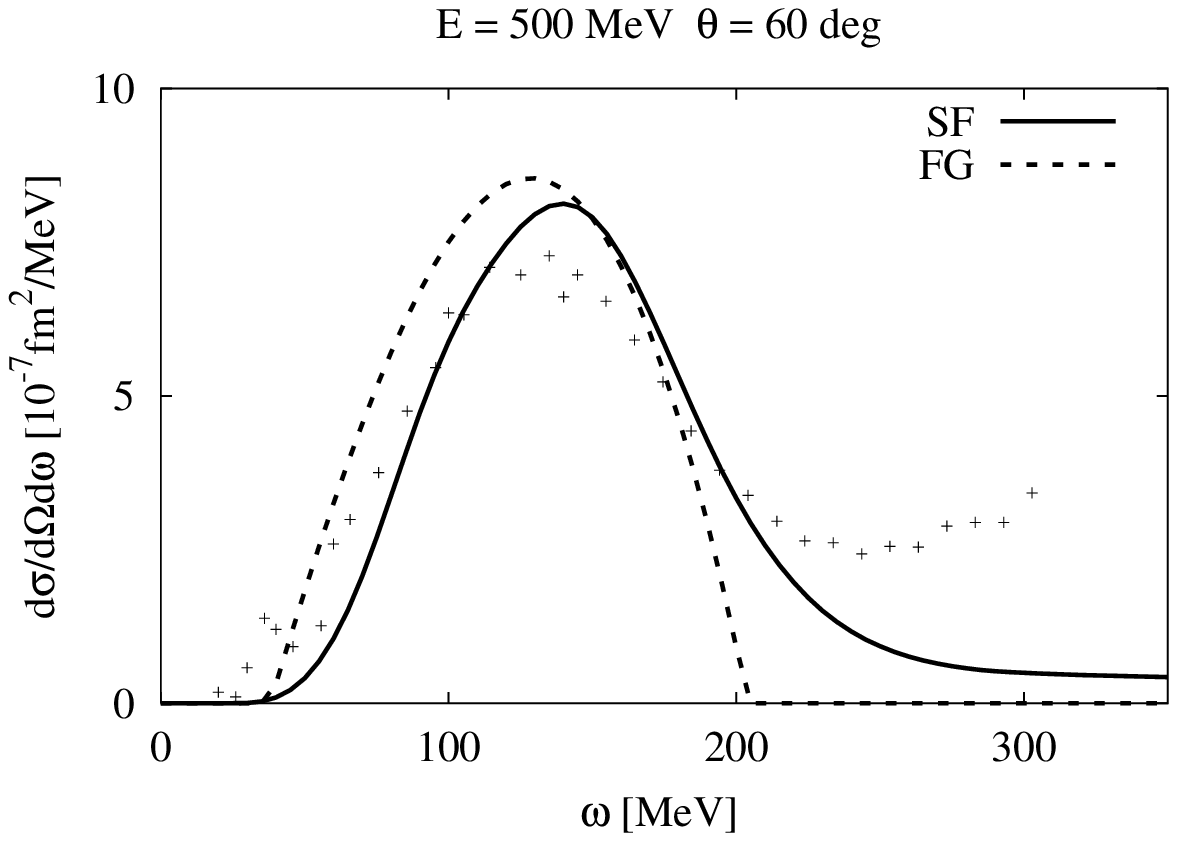}
\includegraphics[scale=0.5]{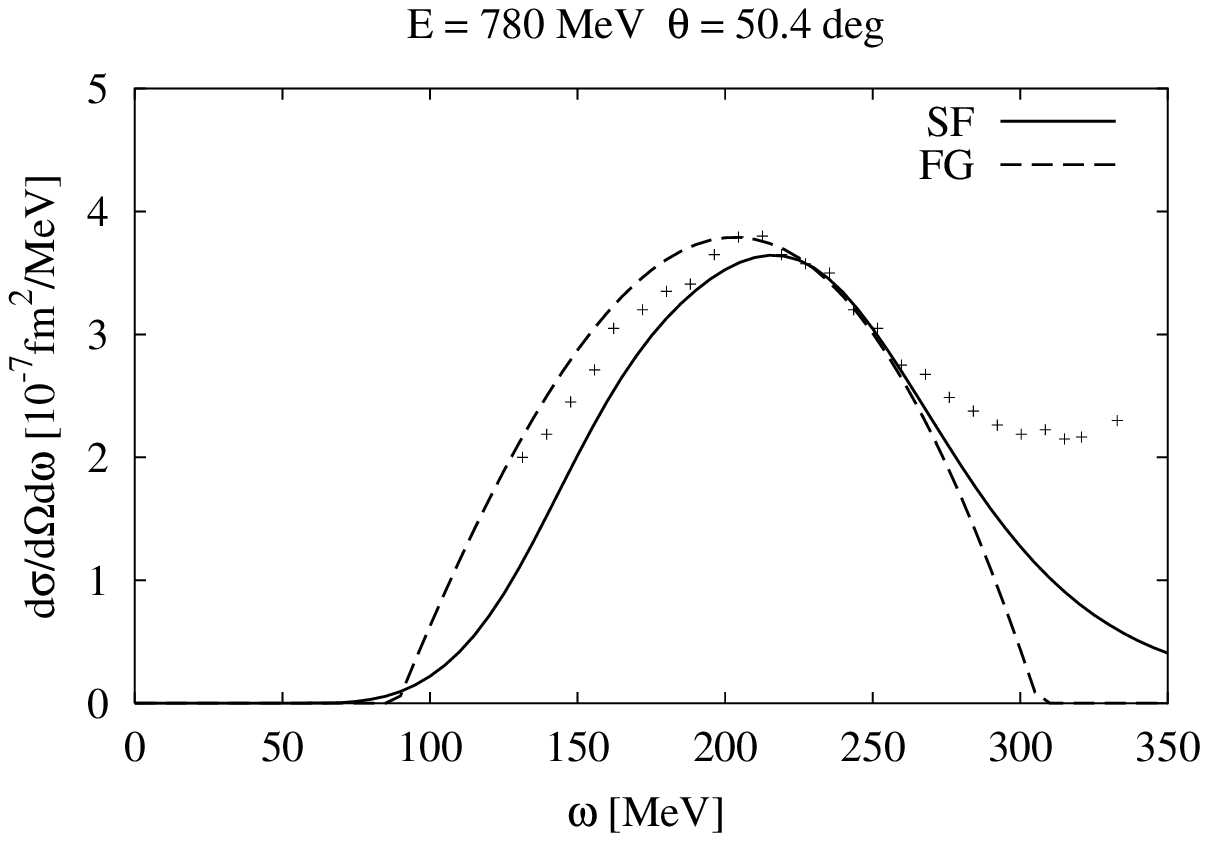}
\end{center}
\caption{${}^{12}$C$(e',e)$ quasi-elastic scattering cross section
using FG model (dotted curves) and SF model (solid curves),
compared with the $(e,e')$ data [21,22].} \label{carbon}
\end{figure}

In Fig.~\ref{carbon} we show a similar calculation for
electron-${}^{12}$C quasi-elastic scattering.  We see the same
trend here.

\begin{figure}[h]
\begin{center}
\includegraphics[scale=0.5]{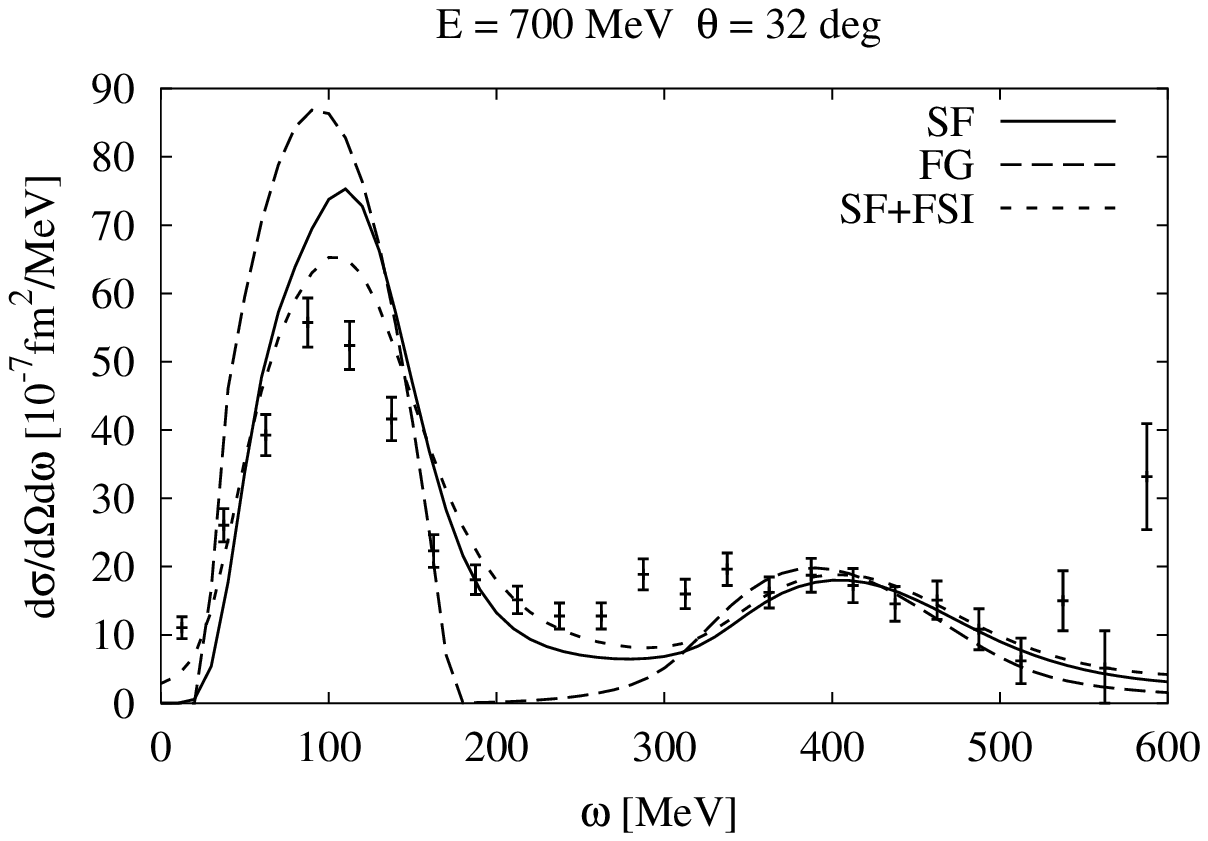}
\includegraphics[scale=0.5]{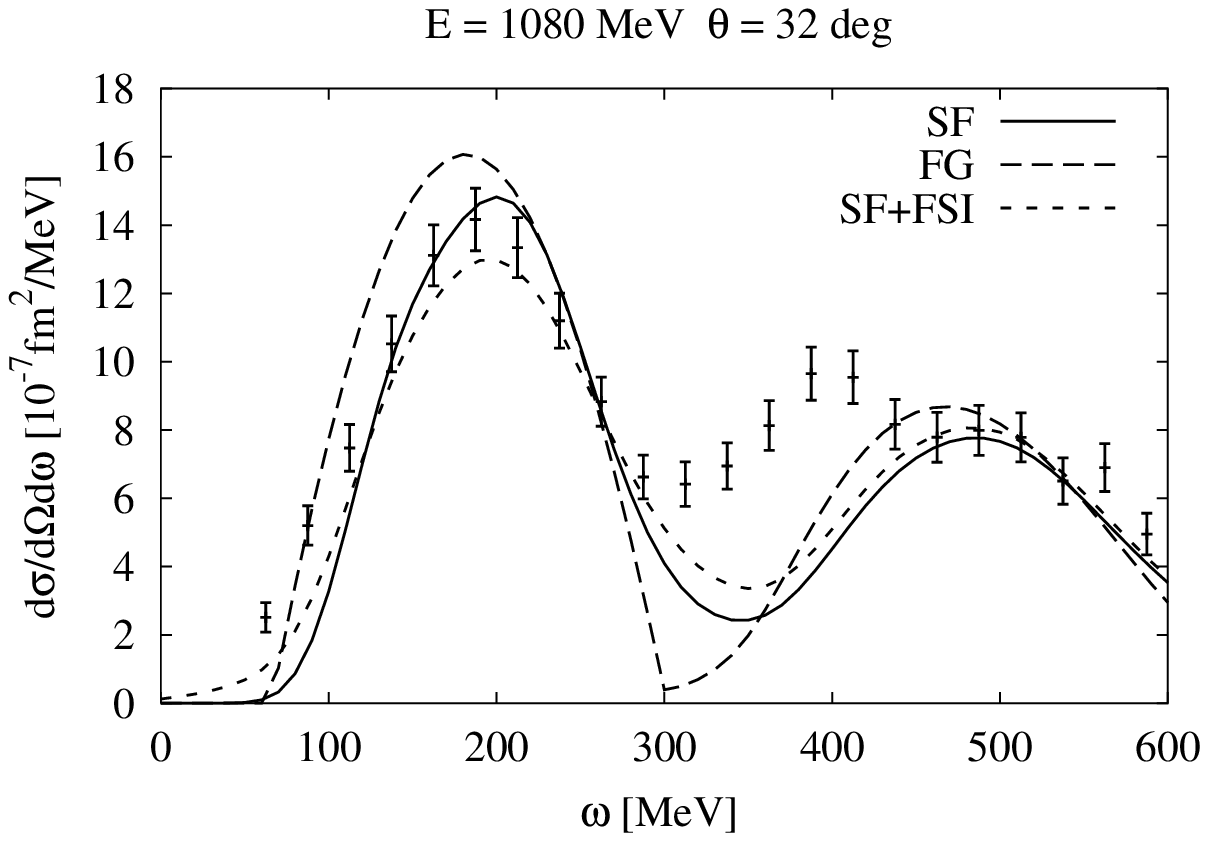}
\end{center}
\caption{Combined cross section of electron-${}^{16}$O$(e',e)$
quasi-elastic scattering and quasi-free $\Delta$ resonance
production.  the figure is the same as Fig.~\ref {oxygen}, except
for SF with the final-state interaction is added and is shown
using the dotted curves.} \label{fsi}
\end{figure}

We also examined the final-state interaction effects using a
simple model based on high-energy nucleon-nucleus optical
potential \cite{pand}. In this, $P_p(\Vp,\omega)$ of SF is changed
as:
\begin{eqnarray}
P_p(\bm p',\omega) &=&\frac{1}{E_p'}\delta(E_p'-\omega)\nonumber\\
&\rightarrow&\frac{1}{E_p'}
  \frac{W/\pi}{(\omega-E_p')^2+W^2/4} \;,
\end{eqnarray}
where the imaginary part of the potential $W$ is related to the
mean-free path of the outgoing nucleon and is written as
\begin{equation}
  W =\frac{1}{2}v\rho\sigma_{NN}
\end{equation}
in terms of the nucleon density $\rho$ and the nucleon-nucleon
cross section $\sigma_{NN}$.  We use $\rho=0.16$ fm${}^{-3}$ and
$\sigma_{NN}=40$ mb for this simple estimate.  Figure \ref{fsi}
shows that the final-state interactions move the strength in the
quasi-elastic peak to the tail regions, and thus broaden the
quasi-elastic peak.  We see that the deficit in the strength
between the two peaks of the quasi-elastic scattering and
quasi-free $\Delta$ production is somewhat reduced and the
agreement with the data is improved somewhat.

In this work, we have not considered other pion production process
than the quasi-free $\Delta$ production.  For this, some
phenomenological \cite{H2model} and theoretical work
\cite{spanish} are available.  We leave this issue and others,
such as exchange-current contributions, as future work.

\subsection{$\nu_\mu$-${}^{16}$O quasi-elastic scattering}

We now turn to the $\nu_\mu$-${}^{16}$O quasi-elastic scattering.

\begin{figure}[h]
\begin{center}
\includegraphics[scale=0.5]{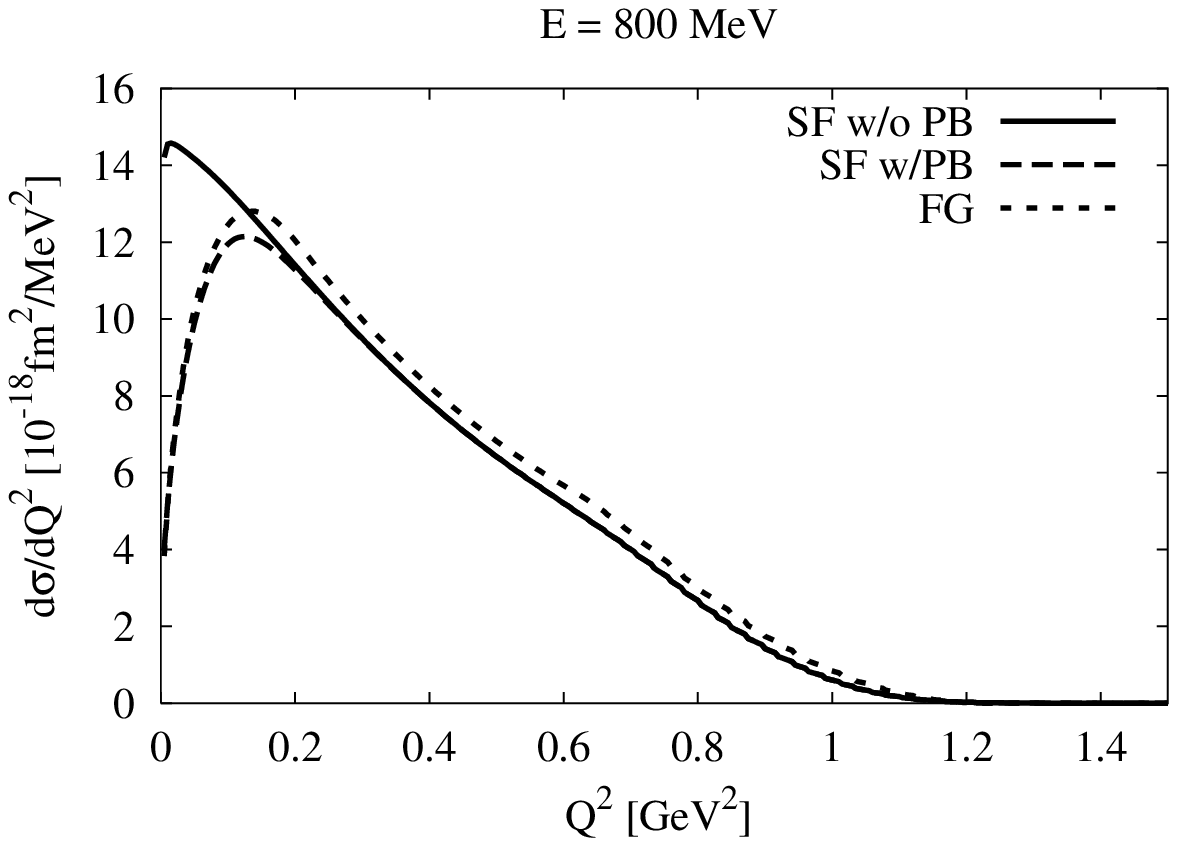}
\includegraphics[scale=0.5]{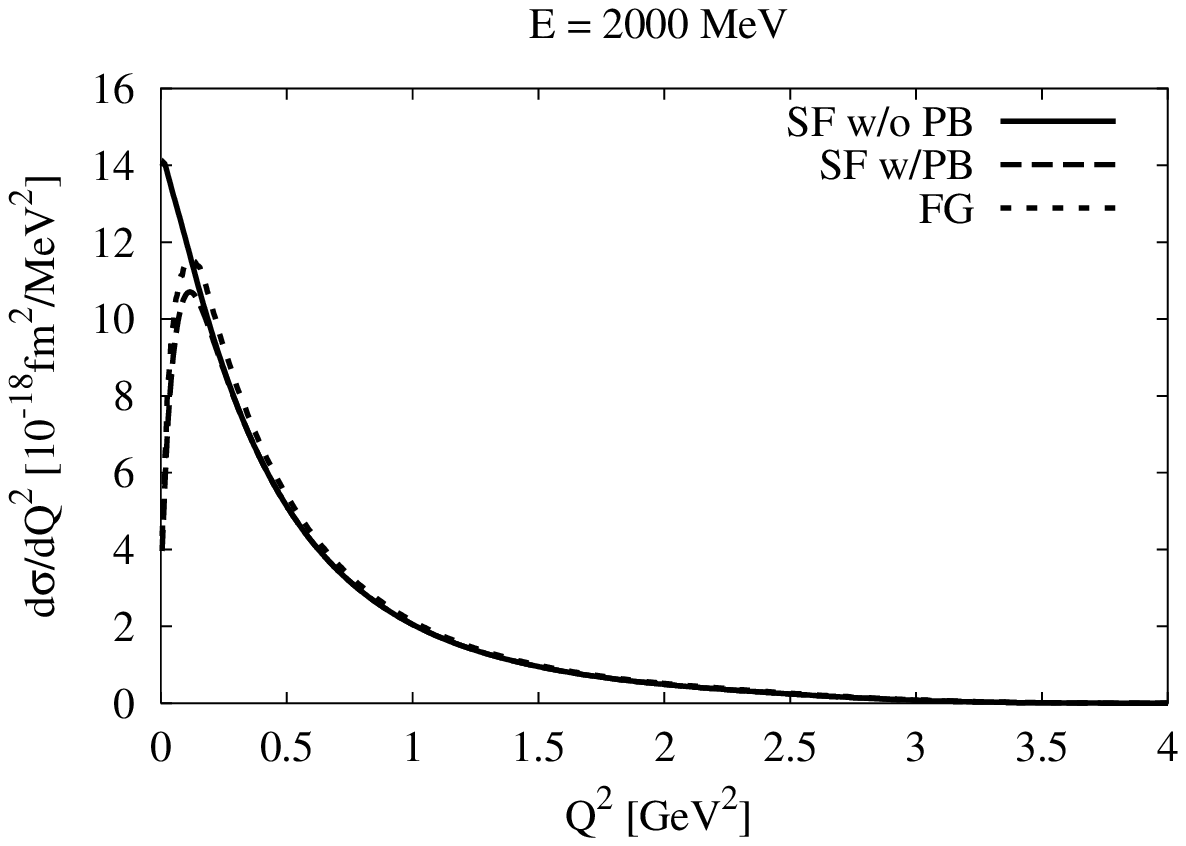}
\end{center}
\caption{Differential cross section of ${}^{16}$O$(\nu_\mu,\mu^-)$
quasi-elastic scattering as a function of square of the momentum
transfer $Q^2$ at the neutrino incident energy $E=0.8$ and 2.0
GeV.  The cross sections are shown, calculated using SF without
and with the Pauli blocking (solid and dash curves, respectively)
and FG (dotted curves.)} \label{q2}
\end{figure}

First, we examine the momentum-transfer $Q^2$ dependence of the
differential cross section using FG and SF. Near the forward, $Q^2$
is small and the Pauli blocking (PB) plays the major role even in
this high-energy region.  Accordingly, in addition to the use of
the same FG as before, we use two-types of SF; without PB as
previously shown in Eq. (6), and with PB as
\begin{eqnarray}
P_p(\bm p',\omega) &=&\frac{1}{E_p'}\delta(E_p'-\omega) \nonumber \\
&\rightarrow& \frac{1}{E_p'}\theta(|p'|-P_F)\delta(E_p'-\omega)
\;.
\end{eqnarray}
In this way, we include in SF a part of final-state interaction
effects, which is generated by PB.  $P_p(\bm p',\omega)$ takes a
more complicated form when the effects are included more
realistically, and the calculation here is a simple estimate.

The $\nu_\mu$-${}^{16}$O quasi-elastic cross section thus
calculated is shown in Fig.~\ref{q2}.  By comparing the cross
sections by the two types of SF cross sections, with and without
PB, we see that the Pauli blocking clearly plays a decisive role
and that the cross sections calculated by SF with
PB and by FG are quite close in the low $Q^2$ region.
This is reasonable since SF with PB and FG use the
same $P_p(\bm p',\omega)$ by incorporating PB in the same way.

Note that the the small $Q^2$ contribution is usually not considered
in the case of electron scattering as the cross section diverges
for $Q^2 \rightarrow 0$ and also the incident beam line
experimentally prevents a measurement close to forward.  In the
case of the high-energy neutrino scattering, the situation is
quite different as we know: Ample events appear near the forward,
making it vital to understand the forward depression as shown in
Fig.~\ref{q2}.
We note that, though not shown here, the electron 
scattering cross sections 
given in Figs.~2-3 are changed little by the inclusion of the Pauli 
Blocking effect (9).  The momentum transfer associated with the 
kinematics examined in these figures is large.

\begin{figure}[h]
\begin{center}
\includegraphics[scale=0.5]{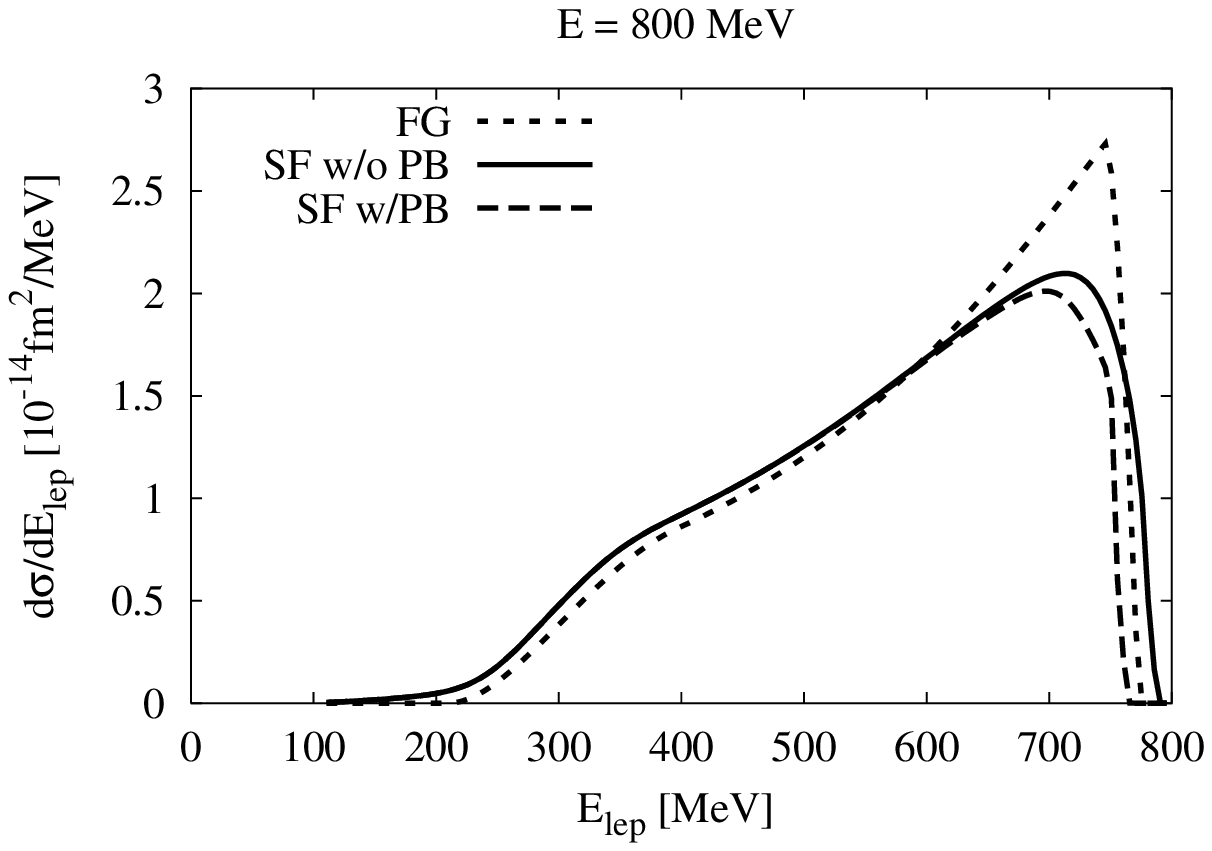}
\includegraphics[scale=0.5]{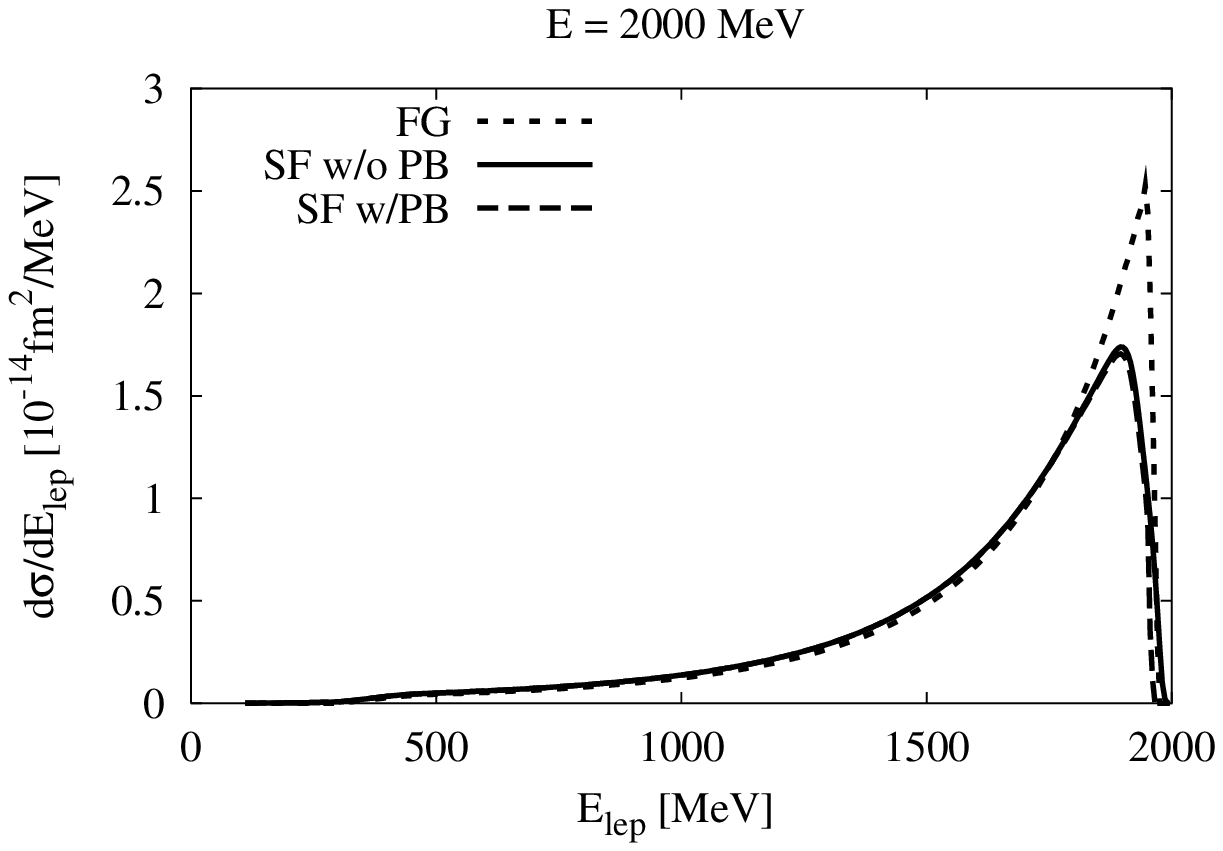}
\end{center}
\caption{Differential cross section of ${}^{16}$O$(\nu_\mu,\mu^-)$
quasi-elastic scattering as a function of the scattered muon
energy for $E=0.8$ and 2.0 GeV, using SF without and with the
Pauli blocking (PB) (solid and dash curves, respectively) and FG
(dotted curves).} \label{del}
\end{figure}

Second, we examine the $\nu_\mu$-nucleus cross sections as a
function of the scattered muon energy.  Figure \ref{del} compares
the cross sections calculated by the three different ways as
before: FG and SF with and without PB.  
The difference between SP with and without PB observed in low $Q^2$ region
in Fig.5 corresponds to that at the higher end of 
$E_{\mu}$ spectrum in this figure. 
FG yields a larger high-energy peak contribution. This
difference is {\it not} due to the Pauli blocking, but due to the
nuclear correlation effects in the spectral function: The
difference is after all associated with scattered muons at a high
energy.  
This difference should show up in the forward angle cross section.
We emphasize that {\it this difference may have a direct
effect on neutrino oscillation measurements}.

\section{Conclusion}
We have carried out a relativistic calculation of the cross
sections of electron-${}^{16}$O and -$^{12}$C quasi-elastic
scattering and quasi-free $\Delta$ production and have compared
with the data systematically in the wide energy range of 0.5 --
1.5 GeV. Using the same formalism, we have then examined the
$\nu_\mu$ quasi-elastic scattering from ${}^{16}$O.

We find that Spectral function calculation agrees better with the
experimental data $A(e,e')$ than a uniform Fermi-gas model. In
particular, this is important to explain the "dip region" between
the quasi-elastic interaction and the delta production kinematics.
A uniform Fermi-gas model cannot fill the dip region. In addition,
the total cross section does not change between the uniform
Fermi-Gas model and the spectral function calculation within 10\%, 
but the differential cross section as functions of the scattered muon
energy changes as much as 20--30\%. We think that this is very
important for the future neutrino experiments aiming at $1\%$ or
better precision.

\section*{Acknowledgement}

 We thank Omar Benhar for providing us his spectral functions of ${}^{12}$C
 and ${}^{16}$O.  This work is supported by 
 the Japan Society for the Promotion of Science through 
 the Grant-In-Aid for Scientific Research (15540299),
 and
 by the US Department of Energy (DE-FG02-87ER40347) at CSUN and by the US
 National Science Foundation (NSF0244899) at Caltech.

\end{document}